# The geometry of cognition: from neural populations to neurotechnology

*Lucas Bayones*[a,*], *Devika Narain*[b,c,*] & *Román Rossi-Pool*[a,d,*]

[a] Instituto de Fisiología Celular – Neurociencias, Universidad Nacional Autónoma de México, 04510 Mexico City, Mexico

[b] Department of Neuroscience, Erasmus MC, Dr. Molewaterplein 40, 3015 GD Rotterdam, The Netherlands

[c] Donders Centre for Neuroscience, Radboud University, Nijmegen, The Netherlands

[d] Centro de Ciencias de la Complejidad, Universidad Nacional Autónoma de México, 04510 Mexico City, Mexico

* Corresponding authors. E-mail: lbayones@ifc.unam.mx ; d.narain@erasmusmc.nl ; romar@ifc.unam.mx

## Abstract

Cognition is implemented by distributed neural populations whose collective activity evolves through structured, low-dimensional state spaces. In this review, we survey the analytical and conceptual tools used to study latent population dynamics in biological neural networks. We begin with single-unit coding, which remains essential for interpreting neuronal selectivity, and then review population-level approaches, including population vectors, cosine similarity, principal component analysis and its mnemonic variants, multivariable linear regression axes, demixed principal component analysis, and single-trial latent-variable models, and distinguish operationally defined population subspaces from neural manifolds with reproducible geometry or topology. We use these methods to discuss sensory coding, working memory, decision-making, temporal processing, and sensorimotor transformations. Two organizing principles are emphasized throughout. First, cognitive variables can be expressed in separable or near-orthogonal population dimensions, reducing interference between sensory, mnemonic, decision-related, and motor signals. Second, such geometric separation does not require anatomically or functionally discrete cell classes; separable population readouts can coexist with a continuum of mixed-selective single-neuron responses within a shared biological substrate. We then consider recurrent neural networks as testbeds for generating and evaluating mechanistic hypotheses about how circuits produce low-dimensional dynamics, and intracortical brain–computer interfaces as a translational domain in which population geometry becomes an operational control signal. Overall, this review argues that population dynamics now provide a common language for linking neurophysiology, computational modelling, and neurotechnology.

**Highlights**

- Step-by-step methods, with equations, for analysing neural population dynamics
- Separable population readouts can emerge from a continuum of mixed-selective neurons
- Subspaces and manifolds are distinguished by explicit, testable criteria
- Temporal processing is implemented by flexible low-dimensional population dynamics
- Population geometry links cortical recordings, RNN models, and intracortical BCIs

## 1. Introduction

Cognitive operations such as perception, working memory, decision-making, temporal processing, and motor planning are not implemented by isolated neurons. They arise from distributed neural populations whose activity changes over time, across task epochs, and across behavioural contexts. Classical neurophysiology approached this problem neuron by neuron: researchers recorded single-unit activity with extracellular electrodes while animals or humans performed controlled behavioural tasks, and used tuning curves to relate firing rates to sensory stimuli, mnemonic variables, choices, or movements. This approach produced foundational insights into somatosensory coding, spatial and parametric working memory, perceptual decision-making, and motor control (Mountcastle et al., 1967; Georgopoulos et al., 1986; Funahashi et al., 1989; Romo et al., 1999; Shadlen and Newsome, 2001; Luna et al., 2005). Single-neuron analyses remain indispensable because they reveal selectivity, latency, reliability, tuning, and the temporal structure of neuronal responses; in many sensory systems, tuning functions are still among the most direct and interpretable descriptions of the neural code. In higher-order cortical and subcortical circuits, however, neurons often display heterogeneous, nonlinear, and multiplexed responses that are difficult to interpret from single-neuron activity alone. For these circuits, population-level analyses provide a complementary description of how task variables are jointly encoded, transformed, and read out.

Two developments motivated a shift from this neuron-by-neuron view to a population-dynamics framework. The first is technical. Modern electrophysiological recordings can sample hundreds or thousands of units simultaneously across one or multiple brain regions, generating datasets whose natural object is not the isolated tuning curve but the joint activity pattern across neurons. This increase in scale has been enabled both by large-scale recording technologies (Jun et al., 2017; Steinmetz et al., 2021) and by increasingly automated spike-sorting pipelines (Pachitariu et al., 2016; Yger et al., 2018; Suárez-Barrera et al., 2025). The second is conceptual. Many neurons in these higher-order circuits display mixed selectivity: their firing rates depend on nonlinear combinations of task variables, and their coding profile can change over the course of a trial (Rigotti et al., 2013; Fusi et al., 2016; Rossi-Pool et al., 2017). Such heterogeneity is not merely noise: it can reflect multiplexed, context-dependent coding whose organization becomes clearer when activity is analysed at the population level. Population analyses can reveal whether apparently heterogeneous single-neuron responses participate in structured collective dynamics, such as low-dimensional trajectories, attractor-like organization, or task-dependent readout axes (Saxena and Cunningham, 2019; Barack and Krakauer, 2021; Ebitz and Hayden, 2021).

Population dynamics occupy an intermediate scale between single spikes and macroscopic signals such as EEG, MEG, or fMRI. The central object is the population state: a vector describing the activity of the recorded neurons at a given moment. As the task unfolds, this vector traces a trajectory through a high-dimensional neural state space. This formulation embeds single-neuron selectivity within a geometric description of collective activity**. The focus thus shifts** toward how task variables are arranged, transformed, separated into axes or subspaces, and read out at the population level. Across stimuli, choices, contexts, or task epochs, these trajectories often occupy a restricted low-dimensional structure within the state space. When this structure is approximately linear and defined operationally by a projection method, it is best described as a population subspace. When it has a reproducible, possibly curved or nonlinear, geometry or topology—whether imposed by the circuit's intrinsic dynamics or inherited from the structure of the variables it encodes—it is more appropriately described as a neural manifold (Gallego et al., 2017; Jazayeri and Afraz, 2017; Chaudhuri et al., 2019; Jazayeri and Ostojic, 2021; Langdon et al., 2023; Perich et al., 2025). Because a manifold is an inferred property of population activity rather than a direct output of a projection method, the two terms are not interchangeable; we return to the criteria that distinguish them in Section 3. Nonlinear manifold-learning methods such as isometric feature mapping (ISOMAP; Tenenbaum et al., 2000), t-distributed stochastic neighbour embedding (t-SNE; van der Maaten and Hinton, 2008), and uniform manifold approximation and projection (UMAP; McInnes et al., 2018), more recent approaches such as RATS (Kohli et al., 2025) and MARBLE (Gosztolai et al., 2025), and representation-learning methods such as CEBRA (Schneider et al., 2023) are often used to recover manifolds and embed them into lower-dimensional spaces. Because these embeddings can distort geometry, their results require validation (Kohli et al., 2025).

This population perspective has reshaped the study of working memory, decision-making, temporal processing, and sensorimotor transformations. In parametric working memory, population analyses showed that information can be maintained even when the single-neuron code changes over time, supporting dynamic population codes that can coexist with a stable mnemonic subspace (Brody et al., 2003; Barak et al., 2010; Machens et al., 2010; Murray et al., 2017). In context-dependent decision-making, task-relevant and task-irrelevant variables can coexist within the same population and be expressed along different readout axes or dynamical regimes (Mante et al., 2013; Aoi et al., 2020). In motor and premotor circuits, preparatory and movement-related activity can occupy distinct subspaces, allowing motor plans to be formed without producing premature movement (Kaufman et al., 2014; Elsayed et al., 2016), while movement control itself can be described within low-dimensional structure, often termed a motor manifold (Gallego et al., 2017; Russo et al., 2018). These examples illustrate related but distinct levels of population analysis: mixed selectivity describes the response properties of individual neurons; latent trajectories describe how population states evolve over time; subspaces define linear readout dimensions; manifolds refer to structured low-dimensional geometry; and attractor-like dynamics describe candidate mechanisms that constrain the evolution of population activity.

Somatosensory, parietal, and premotor circuits are a useful bridge between classical coding and population dynamics. Studies of vibrotactile perception have shown how sensory, mnemonic, and decision-related variables are represented across cortical hierarchies, from early somatosensory areas to secondary somatosensory, parietal, premotor, and prefrontal circuits (Romo and Rossi-Pool, 2020). In this domain, population-level analyses have shown that sensory, categorical, temporal, and decision-related signals can be embedded in shared neural populations while remaining separable at the level of geometry (Rossi-Pool et al., 2017, 2019; Diaz-deLeon et al., 2022; Bayones et al., 2024). Temporal processing illustrates the same logic especially clearly because time is not simply an external stimulus feature but a latent variable that must be generated, maintained, compared, or transformed by the brain. Cortical populations can encode elapsed time, produced intervals, temporal categories, or rhythmic tempo through evolving trajectories, cyclic dynamics, temporal scaling, or low-dimensional attractor-like mechanisms (Wang et al., 2018; Gámez et al., 2019; Beiran et al., 2023; Díaz et al., 2025).

A central theme of this review is that single-neuron heterogeneity and population-level structure are not opposing descriptions, but complementary levels of analysis. Individual neurons in association and premotor cortices often show mixed-selective responses, with coding profiles that vary across task epochs. At the same time, the joint activity of these neurons can reveal low-dimensional structure and separable readout dimensions for sensory, mnemonic, temporal, decision-related, or motor variables (Kaufman et al., 2014; Elsayed et al., 2016; Kobak et al., 2016; Rossi-Pool et al., 2017; Libby and Buschman, 2021; Xie et al., 2022). Population geometry adds an account of how heterogeneous responses can support stable and flexible computations. Separable or near-orthogonal readouts should therefore be interpreted as properties of population organization and task-dependent decoding, not necessarily as evidence for anatomically segregated neuronal classes (Diaz-deLeon et al., 2022; Bayones et al., 2024). This distinction is essential for interpreting association and premotor cortices, where flexible cognition depends on the coexistence of multiple signals within shared neural populations.

The same state-space formalism also links descriptive neurophysiology to computational modelling and neurotechnology. Single-trial and model-based dimensionality-reduction methods, including tensor component analysis (TCA), model-based targeted dimensionality reduction (mTDR), generalized multilinear models (GMLM), and latent factor analysis via dynamical systems (LFADS), recover trial-to-trial structure that is lost in trial averages and relate latent dynamics to behaviour, learning, and variability (Aoi and Pillow, 2018; Pandarinath et al., 2018; Williams et al., 2018; Aoi et al., 2020; Latimer and Freedman, 2023). Recurrent neural networks (RNNs) provide testbeds for generating and evaluating mechanistic hypotheses about how low-dimensional dynamics arise from connectivity, constraints, and task demands (Song et al., 2016; Yang et al., 2019; Yang and Wang, 2020; Dubreuil et al., 2022; Beiran et al., 2023). Finally, brain–computer interfaces (BCIs) are a closed-loop domain in which population geometry becomes an operational control signal and can reveal how the intrinsic low-dimensional structure of population activity constrains, guides, or adapts during learning (Sadtler et al., 2014; Oby et al., 2019). In clinical and translational settings, related decoding approaches have enabled high-performance movement and communication neuroprostheses (Willett et al., 2021, 2023; Metzger et al., 2023; Card et al., 2024). This review follows the progression from neural coding to population geometry, computation, and neurotechnology, covering single-unit and population-level methods, demixing, orthogonality and separability, continuity of single-neuron responses, temporal processing, single-trial dynamics, artificial neural networks, and intracortical brain–computer interfaces.

## 2. From spikes to single-unit coding

During cognitive tasks, neuronal activity is often recorded from one or more brain regions using extracellular electrodes that measure voltage fluctuations generated by nearby neural populations. These signals allow analyses of both spiking activity and local field potentials (LFPs). This review focuses on spiking responses; for LFP analysis, see specialized literature (Cohen, 2014). To derive spike trains from extracellular recordings, the broadband signal is high-pass filtered, thresholded, and segmented into candidate spike waveforms (Fig. 1A). The resulting multi-unit activity reflects the collective spiking responses of neurons close to the electrode. Subsequent waveform sorting attempts to assign detected spikes to putative single neurons (Quiroga et al., 2004).

Standard spike-sorting techniques use waveform features obtained from feature-extraction or dimensionality-reduction methods such as principal component analysis (PCA) or wavelets (Fig. 1A), supplemented by descriptors such as energy or peak-to-peak amplitude (Quiroga et al., 2004). These procedures often require manual curation to refine spike detection and cluster boundaries. More recent approaches, including Kilosort (Pachitariu et al., 2016, 2024) and specialized toolboxes for large electrode arrays (Yger et al., 2018), have substantially improved the speed and scalability of spike sorting.

Nonlinear dimensionality-reduction methods can also be used to visualize and separate spike waveforms. UMAP, for example, embeds high-dimensional data into a low-dimensional space while preserving local structure (McInnes et al., 2018). Applied to extracellular recordings, UMAP-based spike sorting can improve the separation of neuronal units, including sparsely firing neurons, without relying exclusively on manually selected linear waveform features (Fig. 1A; Suárez-Barrera et al., 2025).

After spike sorting, neuronal firing rates are typically estimated by convolving each spike train with a square or Gaussian kernel and then averaging responses across trials or conditions (Dayan and Abbott, 2005). We illustrate these analyses with the temporal pattern discrimination task (TPDT), in which the animal reports whether the temporal structure of a second vibrotactile stimulus matches that of the first (Fig. 1C; Rossi-Pool et al., 2016). Dorsal premotor cortex (DPC) neurons recorded during this task display markedly diverse single-unit coding profiles (Fig. 1D; Rossi-Pool et al., 2017). Several analytical tools can then be used to quantify the relationship between neuronal activity and task variables. Shannon's mutual information (Eq. 1) quantifies, in bits, the statistical dependence between a neuron's response (r) and a task parameter (p) (Panzeri and Treves, 1996; Koren et al., 2023).

$$I = \sum_{r,p} P\,(p)P(r|p)log_2\left(\frac{P(r|p)}{P(r)}\right) \qquad \textit{(Eq. 1)}$$

Mutual information captures arbitrary, not only linear, dependencies between firing rate and task variables (Fig. 1E). It can therefore be used to identify informative neurons, task epochs, or variables that may later enter population-level analyses (Panzeri et al., 2007; Kayser et al., 2010; Rossi-Pool et al., 2016). A complementary nonparametric measure is the area under the receiver operating characteristic curve (AUROC), which compares the response distributions obtained under two conditions (Britten et al., 1996). Depending on the comparison, AUROC can quantify stimulus selectivity, choice-related activity, or differences between correct and error trials. In decision tasks, it is often used to estimate how well a single neuron's firing rate discriminates the subject's report or behavioural outcome (de Lafuente and Romo, 2005; van Vugt et al., 2018).

A third approach is multivariable linear regression, which models firing rate as a weighted combination of task parameters (Eqs. 2–3), including stimulus features, memory variables, choices, context, and trial history (Brody et al., 2003; Mante et al., 2013; Carnevale et al., 2015; Akrami et al., 2018).

$$r(t) = \beta_0(t) + \beta_1(t)p \qquad \textit{(Eq. 2)}$$

$$r(t) = \beta_0(t) + \beta_1(t)p_1 + \beta_2(t)p_2 \qquad \textit{(Eq. 3)}$$

In frontal cortex, regression analyses have revealed neurons whose stimulus-related activity peaks early, is sustained through the delay, or emerges late in the trial. These profiles suggest distinct contributions to information encoding, maintenance, and retrieval (Brody et al., 2003; Murray et al., 2017). The same framework can be extended to include the influence of previous trials on current neural responses and behaviour (Akrami et al., 2018). More broadly, single-unit analyses have been central to characterizing mixed selectivity, in which neurons encode multiple task parameters simultaneously or switch their coding profile across time within a trial (Rigotti et al., 2013; Fusi et al., 2016). This heterogeneity has been documented extensively in prefrontal, frontoparietal, and premotor circuits (Chaisangmongkon et al., 2017; Rossi-Pool et al., 2017), and motivates the transition from neuron-by-neuron descriptions to the population-level methods developed in the next section.

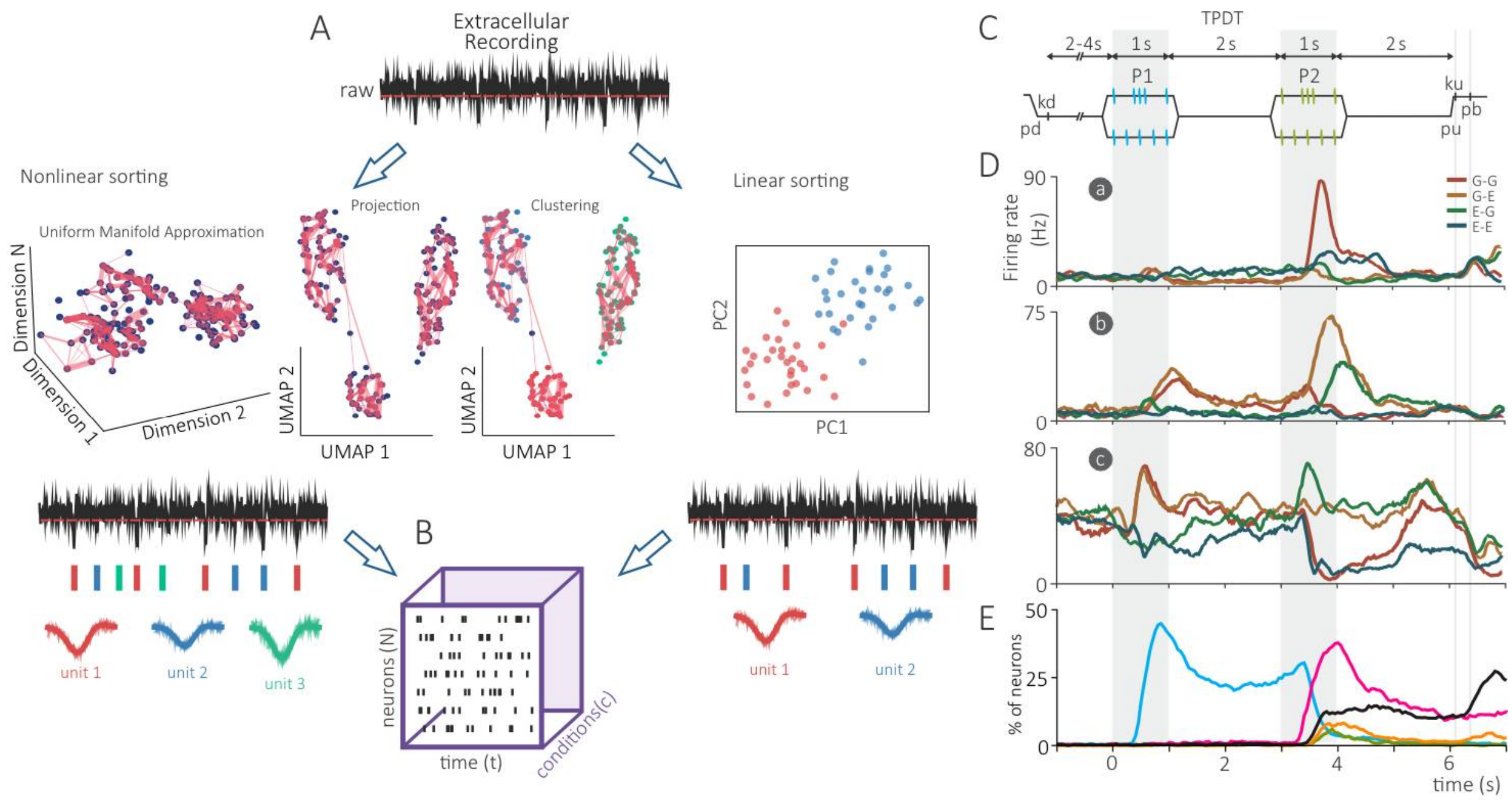

**Figure 1. Single-unit analysis: from spikes to coding dynamics.** (A) Extracellular recordings are high-pass filtered, candidate spikes are detected, and waveform features are clustered to assign events to putative single units. Clustering can be performed in a linear or nonlinear reduced-dimensionality space, such as PC1–PC2 or a UMAP embedding. (B) Recorded activity can be organized as a high-dimensional tensor with axes corresponding to neurons (N), time (t), and conditions (c). (C) Temporal pattern discrimination task (TPDT): the animal reports whether the temporal structure of the second stimulus (P2) matches that of the first stimulus (P1), yielding four condition classes: G–G, G–E, E–G, and E–E (G, grouped; E, extended patterns). Each stimulus lasts 1 s and is followed by a 2-s delay; pd, probe down; kd, key down; pu, probe up; ku, key up; pb, push button. (D) Three example dorsal premotor cortex (DPC) neurons (a–c), with firing rates averaged by class, illustrate the diversity of single-unit coding. (E) Percentage of DPC neurons with significant coding of P1 pattern (cyan), P2 pattern (green), class (pink), partial decision (orange), and complete decision (black) as a function of time; P1 coding persists until mid-P2, when class coding emerges and contributes to the decision signal. Adapted from Rossi-Pool et al. (2016) and Suárez-Barrera et al. (2025).

## 3. Population geometry: vectors, similarity, and PCA

Population-level analyses begin from a simple question: how similar are the activity states of a neural population across conditions, and how do these states evolve over time? To address this question, the activity of the recorded neurons can be organized as a tensor with axes corresponding to neurons (N), time (t), and conditions (c) (Fig. 1B). At each time point, the population response under condition c can be represented as an N-dimensional vector of trial-averaged firing rates, where each entry corresponds to one recorded neuron (Eq. 4):

$$\bar{r}_c(t) = [\bar{r}_c^1(t), \bar{r}_c^2(t), \dots, \bar{r}_c^N(t)] \qquad \textit{(Eq. 4)}$$

This representation makes it possible to compare neural activity geometrically. One simple measure is cosine similarity, which quantifies the angle between two population states, for example those of conditions *i* and *j* at time *t* (Eq. 5):

$$\cos\theta(t) = \frac{\bar{r}_i(t) \cdot \bar{r}_j(t)}{|\bar{r}_i(t)||\bar{r}_j(t)|} \qquad \textit{(Eq. 5)}$$

Cosine similarity can be used to compare population states across conditions at a fixed time, or across time points within the same condition. In an early application to prefrontal cortex activity during a vibrotactile discrimination task, Barak and colleagues used population-state similarity to show that frequency information is preserved across the delay even though the state that carries it evolves continuously (Fig. 2A–C; Barak et al., 2010; Machens et al., 2010). This result anticipated a now-standard conclusion: working memory can be implemented by dynamic population codes, which can coexist with a stable mnemonic subspace (Murray et al., 2017). Whereas cosine similarity compares population states pairwise, PCA characterizes the geometry of the entire set of states by projecting activity onto orthogonal axes that capture progressively smaller fractions of variance (Eqs. 6–8).

$$C_{ij} = \frac{1}{T\,N_c}\sum_{c=1}^{N_c}\sum_{t=1}^{T}\left(r_c^i(t) - \bar{r}^i\right)\left(r_c^j(t) - \bar{r}^j\right) \qquad \textit{(Eq. 6)}$$

$$C = Q\Lambda Q^T \qquad \textit{(Eq. 7)}$$

$$PC_k(t,c) = \sum_{i=1}^{N} q_i^k\left(r_c^i(t) - \bar{r}^i\right) \qquad \textit{(Eq. 8)}$$

Here, $C$ is the $N \times N$ covariance matrix of trial-averaged responses, computed across time points and conditions (Eq. 6), where $\bar{r}^i$ is the mean response of neuron i across all time points and conditions and $N_c$ is the number of conditions. Its eigendecomposition (Eq. 7) yields orthogonal principal axes (the columns of $Q$) ranked by their explained variance (the eigenvalues in $\Lambda$). Projections onto the leading axes (Eq. 8), where $q_i^k$ is the $i$-th element of the $k$-th column of $Q$, summarize the dominant modes of population activity.

Because neural populations exhibit correlated activity and coordinated task-dependent modulations, their responses typically occupy a low-dimensional structure within the full N-dimensional state space (Fig. 2D; Rossi-Pool and Romo, 2019). In linear analyses such as PCA, this structure is described as a population subspace: a set of orthogonal axes that captures the dominant modes of population variance. The term neural manifold refers to something more specific: the reproducible geometry and topology of the set of states a circuit actually occupies, which may be curved or nonlinear and therefore only approximated by a linear subspace (Gallego et al., 2017; Langdon et al., 2023; Perich et al., 2025). This structure can arise from two sources that are not mutually exclusive. In one, intrinsic circuit dynamics constrain the trajectories a population can follow, so that the manifold is traced out by the temporal evolution of the state, as in motor preparation and execution (Elsayed et al., 2016; Russo et al., 2018). In the other, the manifold is inherited from the structure of the variables the population encodes (external, such as head direction or spatial position, or internal, such as a remembered stimulus value), so that its topology mirrors that of the encoded space: a ring for circular variables (Chaudhuri et al., 2019), a torus for the periodic lattice of grid cells (Gardner et al., 2022), a line for a one-dimensional parametric memory (Fig. 2F, G). Under this second view, the population state is a point on the manifold indexed by the encoded variable, and trajectories arise only when that variable changes. The two views converge in attractor networks, where intrinsic dynamics maintain a manifold whose topology matches the encoded variable and which persists even in the absence of sensory input (Chaudhuri et al., 2019; Langdon et al., 2023). PCA and related linear methods can reveal a low-dimensional subspace in either case, but additional evidence is needed before that subspace is interpreted as a neural manifold: reproducibility across conditions and sessions, preserved topology under nonlinear embedding (Kohli et al., 2025), and a mechanistic account of the constraint.

In PCA, leading components can capture large condition-independent changes in activity, including trial progression, whereas task-relevant variables may be represented in other components or subspaces (Fig. 2E; Machens et al., 2010; Kaufman et al., 2016; Rossi-Pool et al., 2017, 2019; Cueva et al., 2020). The distinction between condition-independent and task-dependent components becomes explicit in demixed PCA (Kobak et al., 2016). A practical consequence of this population-level view is that the relevant low-dimensional dynamics may be robust to the precise identity of sorted units. Indeed, multi-unit activity can often recover population dynamics similar to those obtained from sorted single-unit activity (Trautmann et al., 2019).

Several PCA-based strategies have been used to emphasize task-relevant structure within population dynamics. One strategy computes the covariance matrix within a restricted time window (Eq. 9), so that the resulting axes capture task-variable coding expressed during a particular epoch. These components, however, remain tied to the chosen window and can conflate variable-related and temporal structure (Murray et al., 2017; Rossi-Pool et al., 2017). A second approach averages firing rates over the delay period before computing the covariance matrix, yielding mnemonic principal components that emphasize stable working-memory structure (Eqs. 10–13; Murray et al., 2017):

$$C_{ij}(t_1, t_2) = \frac{1}{T_w N_c} \sum_{c=1}^{N_c} \sum_{t=t_1}^{t_2} \left(r_c^i(t) - \bar{r}^i\right)\left(r_c^j(t) - \bar{r}^j\right) \qquad \textit{(Eq. 9)}$$

$$\left\langle r_c^i \right\rangle = \frac{1}{T_d} \sum_{t \in D} r_c^i\,(t) \qquad \textit{(Eq. 10)}$$

$$\langle \bar{r}_c \rangle = [\langle r_c^1 \rangle, \langle r_c^2 \rangle, \dots, \langle r_c^N \rangle] \qquad \textit{(Eq. 11)}$$

$$C_{ij} = \frac{1}{N_c} \sum_{c=1}^{N_c} \left(\left\langle r_c^i \right\rangle - \left\langle \bar{r}^i \right\rangle\right)\left(\left\langle r_c^j \right\rangle - \left\langle \bar{r}^j \right\rangle\right) \qquad \textit{(Eq. 12)}$$

$$mPC_k(c) = \sum_{i=1}^{N} q_i^k \left(\langle r_c^i \rangle - \langle \bar{r}^i \rangle\right) \quad \text{(Eq. 13)}$$

Here, $T_w$ is the number of time points in the window $[t_1, t_2]$ (Eq. 9), and $\langle r_c^i \rangle$ denotes the delay-averaged firing rate of neuron $i$ in condition $c$, computed over the $T_d$ time points of the delay period $D$ (Eqs. 10–11), and $\langle \bar{r}^i \rangle$ is its mean across conditions. The covariance of these condition-wise averages (Eq. 12) yields mnemonic principal components onto which each condition is projected (Eq. 13). The resulting mnemonic PCs (mPCs) reveal low-dimensional attractor-like organization underlying parametric working memory: a line-like geometry for frequency memory and a ring-like geometry for spatial location (Fig. 2F, G). However, neither time-windowed PCA nor mnemonic PCA explicitly demixes temporal and task-dependent components. Section 4 introduces methods designed to separate these contributions more directly.

***When is a subspace a manifold?*** Operationally, low-dimensional structure identified by linear methods is best described as a subspace. The stronger term manifold is warranted when additional criteria are met: the geometry is reproducible across sessions, animals, or data splits; it exhibits nonlinear or topological structure that is not captured by linear projections, such as the ring-like organization of remembered spatial location (Fig. 2G); and it demonstrably constrains population activity, either the trajectories the circuit can generate or the mapping from encoded variables to neural states (Gallego et al., 2017; Jazayeri and Ostojic, 2021; Langdon et al., 2023; Perich et al., 2025). Throughout this review, we use subspace when structure is defined operationally by a linear method, and reserve manifold for cases in which such additional evidence exists.

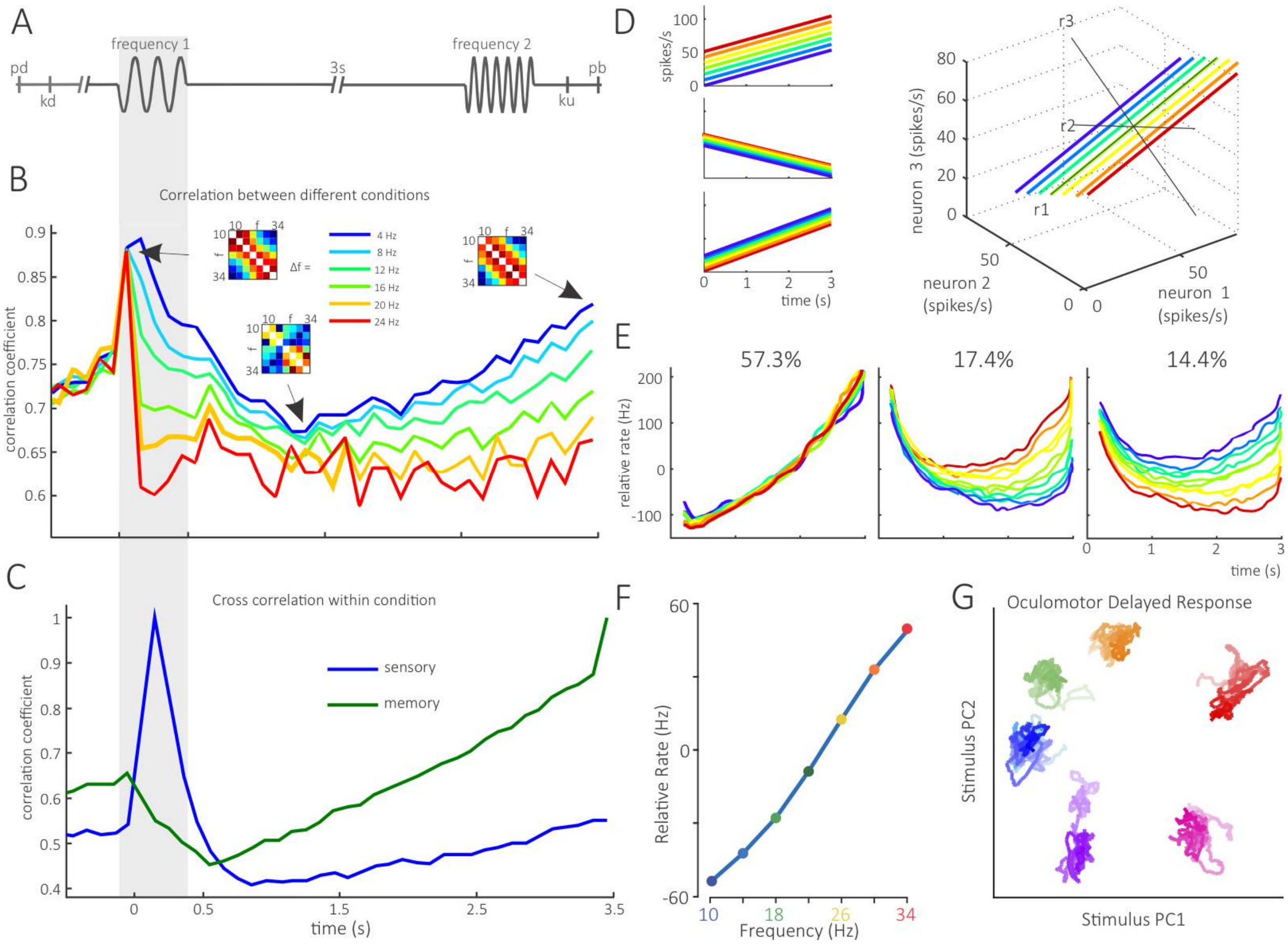

**Figure 2. Low-dimensional population dynamics during parametric working memory.** (A) Vibrotactile frequency-discrimination task (Romo et al., 1999): two flutter stimuli (f1, f2) are delivered to the fingertip and separated by a ~3 s delay; the monkey reports whether f1 > f2 (pd/kd, probe/key down; ku, key up; pb, push button). (B) Across-condition correlations of the population state over time, grouped by frequency difference Δf (4–24 Hz): conditions initially converge to a common stimulus-driven state and later separate by Δf, indicating a parametric population code for frequency. Insets show pairwise correlation matrices at selected times. (C) Within-condition cross-correlations referenced to a sensory state and a late-delay memory state: the weak correlation between reference states reveals distinct sensory and mnemonic population representations of the same frequency. (D) Dimensionality-reduction schematic: synthetic firing rates from three example neurons define a population trajectory confined to a low-dimensional plane within neural state space. (E) Leading principal components of prefrontal cortex (PFC) activity during the delay show that a few dimensions capture most firing-rate variance, with the leading component largely reflecting condition-independent temporal modulation. (F) In the frequency-discrimination task, single neurons show monotonic tuning to stimulus frequency, and the population defines a low-dimensional stimulus subspace, with line-like geometry, in which remembered frequency is stably represented. (G) A similar low-dimensional mnemonic organization arises in an oculomotor delayed-response task, where the stored variable is spatial location and the mnemonic subspace shows a ring-like geometry. Panels B, C adapted from Barak et al. (2010); D, E from Machens et al. (2010); F, G from Murray et al. (2017), with permission of the authors.

## 4. Demixing and orthogonality

PCA identifies orthogonal axes of maximal population variance, but these axes are not tied to individual task variables and can combine temporal, sensory, mnemonic, and decision-related signals. Multivariable linear regression is a complementary way to define variable-specific axes. In this framework, the firing rate of each neuron is modelled as a weighted sum of task variables, or regressors, such as stimulus features, context, choice, or behavioural report (Eq. 14). For neuron $i$, the coefficient $\beta_k^i(t)$ quantifies how strongly its firing rate at time $t$ is modulated by variable $x_k$.

$$r^i(t) = \beta_0^i(t) + \sum_k \beta_k^i(t) x_k \qquad \textit{(Eq. 14)}$$

Because regression coefficients vary over time, a projection axis must be defined at a selected reference time point or task epoch, denoted here as $t^*$. For a given variable $x_k$, the coefficients estimated at $t^*$ are collected across neurons to form a population vector (Eq. 15):

$$a_k(t^*) = [\beta_k^1(t^*), \beta_k^2(t^*), \dots, \beta_k^N(t^*)] \qquad \textit{(Eq. 15)}$$

This vector is then normalized to obtain a unit-norm projection axis, which remains fixed and can be used to project population activity over time (Eq. 16):

$$\hat{a}_k(t^*) = \frac{a_k(t^*)}{|a_k(t^*)|} = \left[\frac{\beta_k^1(t^*)}{|a_k(t^*)|}, \frac{\beta_k^2(t^*)}{|a_k(t^*)|}, \dots, \frac{\beta_k^N(t^*)}{|a_k(t^*)|}\right] \qquad \textit{(Eq. 16)}$$

Mante et al. (2013) used this approach in a context-dependent selection task to show that prefrontal cortex encodes choice, motion evidence, and colour evidence along distinct population axes, while task context determines which sensory variable is used to guide the decision (Fig. 3A, Bi, Bii). Reverse-engineering a recurrent network trained on the same task showed that this selection is implemented by context-dependent recurrent dynamics: the motion and colour inputs keep the same directions in both contexts, whereas the selection vector aligns with the relevant input, which is then integrated along a line attractor (Fig. 3C). Related regression-based state-space analyses have been applied to detection and delayed-response tasks, where they revealed attractor-like dynamics that gate behavioural reports (Carnevale et al., 2015; Inagaki et al., 2019). However, regression-defined axes do not by themselves provide a full decomposition of population variance across time, task variables, and their interactions. Demixed principal component analysis (dPCA) addresses this limitation by decomposing population activity into marginalizations associated with condition-independent activity, task variables, and variable–time interactions, and then applying reduced-rank regression to obtain encoder–decoder pairs for each component (Eqs. 17–19; Kobak et al., 2016).

$$X_0^i = \frac{1}{T\,N_c} \sum_{c=1}^{N_c} \sum_{t=1}^{T} r_c^i(t) \qquad \text{(Eq. 17)}$$

$$\tilde{X} = X - X_0 \qquad \text{(Eq. 18)}$$

$$\begin{aligned} X_t &= \bar{r}(t) - \bar{r}; \\ X_{t,c} &= \bar{r}(t,c) - \bar{r}(t); \\ X_{t,p_1} &= \bar{r}(t,p_1) - \bar{r}(t); \\ X_{t,p_2} &= \bar{r}(t,p_2) - \bar{r}(t); \\ X_{t,d} &= \bar{r}(t,d) - \bar{r}(t); \end{aligned} \qquad \text{(Eq. 19)}$$

Here, $X$ is the $N \times (T\,N_c)$ matrix of trial-averaged responses, $X_0$contains the mean firing rate of each neuron across time points and conditions (Eq. 17), and $\tilde{X}$ is the mean-centred data (Eq. 18). Each marginalization (Eq. 19) isolates a distinct source of variance: a condition-independent temporal component, components tied to individual task variables (for example, stimulus parameters or the decision), and interaction terms that capture variable-dependent temporal structure. Applied to premotor activity during the TPDT, dPCA separated a dominant condition-independent signal from P1- and class-related components, and revealed that P1 information was distributed across two nearly orthogonal subspaces: one persistent and one transitory, carrying distinct aspects of the same stimulus parameter. The class-related components, in turn, revealed a transition from P1 coding to decision coding (Fig. 3D, E; Rossi-Pool et al., 2017). Related model-based approaches, such as model-based targeted dimensionality reduction (mTDR) and generalized multilinear models (GMLM), extend this logic by embedding dimensionality reduction within probabilistic regression models. Instead of first decomposing trial-averaged activity and then interpreting the resulting components, these methods estimate low-dimensional activity patterns directly as functions of specified task or behavioural covariates. This makes them useful for richer task designs, single-trial analyses, and datasets in which choices, stimuli, trial history, or continuous behavioural variables must be modelled jointly (Aoi and Pillow, 2018; Aoi et al., 2020; Latimer and Freedman, 2023). By contrast, sparse component analysis (SCA) requires no task labels: it identifies temporally sparse latent factors, constrained to occupy orthogonal dimensions, that can recover distinct computations within population activity (Zimnik et al., 2026).

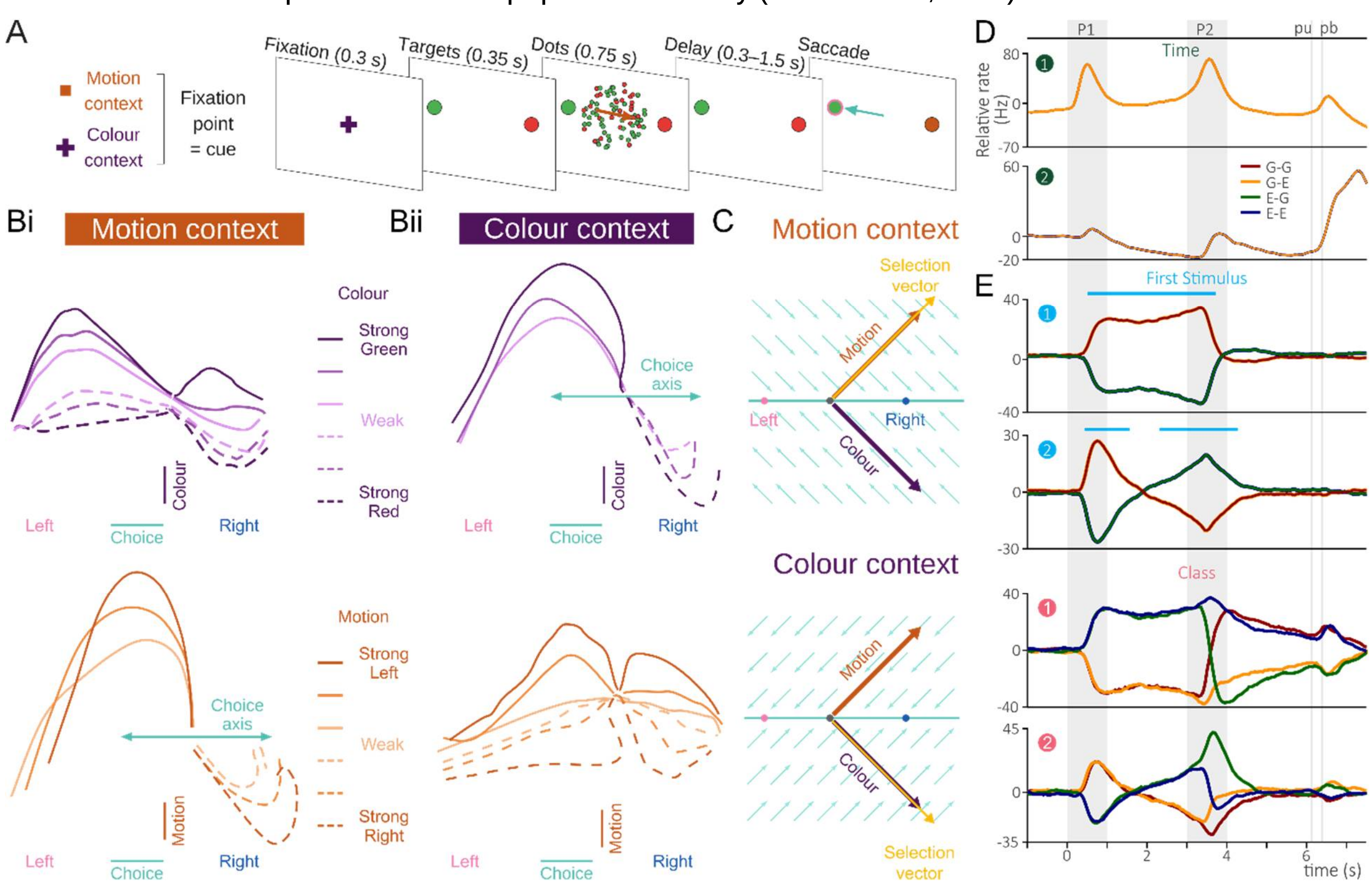

**Figure 3. Regression axes and demixing of population activity.** (A) Context-dependent decision-making task: after fixation and target onset, a random-dot stimulus containing both motion and colour evidence is presented for 0.75 s, followed by a 0.3–1.5 s delay and a saccadic report; a contextual cue instructs the animal to judge either motion or colour while ignoring the other feature. (Bi, Bii) State-space schematics of population trajectories in the motion (Bi) and colour (Bii) contexts, sorted by colour (top) and motion (bottom) evidence: relevant evidence is integrated along the choice axis, whereas irrelevant evidence is represented but not accumulated. (C) Dynamical system implementing the selection vector in the motion (top) and colour (bottom) contexts. The motion and colour inputs keep the same directions in both contexts, whereas context changes the recurrent dynamics so that the selection vector aligns with the relevant input, which is then integrated along the choice axis (line

attractor). (D, E) dPCA of DPC activity during the TPDT separates a dominant condition-independent component from stimulus- and class-related components, and reveals persistent and transitory P1 components, whereas class components reveal a transition from P1 to decision coding, across classes G–G, G–E, E–G, and E–E. Panels A–C are schematics based on Mante et al. (2013); D, E adapted from Rossi-Pool et al. (2017).

***Orthogonality and separability as computational principles.*** Across tasks and areas, a recurring motif is that different task variables can be expressed in separable population dimensions or subspaces. This interpretation requires some care. In PCA, orthogonality is imposed by construction, because components are constrained to be mutually orthogonal axes of maximal variance. By contrast, demixed PCA separates population activity according to task-variable marginalizations, but does not by itself require the subspaces associated with different variables to be orthogonal. Thus, weak alignment or near-orthogonality between regression axes, demixed subspaces, decoder subspaces, or task-specific population modes supports the interpretation of functional separability rather than reflecting a trivial consequence of the analysis method, especially when these are defined or validated across independent epochs, conditions, or data splits. Examples include sensory and mnemonic coding in prefrontal and premotor circuits (Rossi-Pool et al., 2017; Libby and Buschman, 2021), the separation of sensory and delay-period maintenance dynamics in premotor cortex during bimodal detection (Andrade-Ortega et al., 2025), motor preparation and execution in motor cortex (Kaufman et al., 2014; Elsayed et al., 2016), sensory and decision-related signals in premotor circuits (Diaz-deLeon et al., 2022), sequence positions in PFC (Xie et al., 2022), reward- and confidence-related components in orbitofrontal cortex (Kobak et al., 2016), evidence-accumulation and decision-commitment modes during perceptual decisions (Luo et al., 2025), and faithful versus transformed sensory codes in secondary somatosensory cortex (S2) (Rossi-Pool et al., 2021; Bayones et al., 2024). The shared logic is functional separability: a single population can multiplex several variables while reducing crosstalk between readouts. Such separability should not be interpreted as evidence for anatomically segregated or functionally discrete cell classes. A shared population can contain mixed-selective neurons while still supporting separable, and in some cases nearly orthogonal, readouts for sensory, mnemonic, decision-related, or motor variables. The next section develops this point in more detail.

## 5. Continuity of neural responses

The discussion of separability and orthogonality raises a further question: do separable or near-orthogonal population projections imply separate groups of neurons? Do different cognitive variables require distinct neural substrates, or can they arise from the same population through projections onto different axes? The answer affects how cortical circuitry is conceptualized and how single-neuron responses are classified into putative functional types.

Evidence from population recordings suggests that separable, and in some cases nearly orthogonal, population dynamics can emerge from a shared and continuous neural substrate. Bayones et al. (2024) applied demixed PCA to S2 activity during the TPDT and showed that faithful sensory, transitory, and transformed categorical responses were expressed in distinct, nearly orthogonal population subspaces when the data were marginalized with respect to P1 and class (Fig. 4A, B, for the P1 marginalization). To examine how these population-level modes were supported at the single-neuron level, neurons were embedded according to their response profiles using UMAP (McInnes et al., 2018), a nonlinear dimensionality-reduction method that has been used to visualize whether high-dimensional neuronal features form discrete clusters, overlapping classes, or graded continua (Diaz-deLeon et al., 2022; Parra et al., 2022; Feyerabend et al., 2026). Across stable parameter choices, these functional modes did not form discrete clusters but occupied a continuous region of the embedding space. Because low-dimensional embeddings can distort geometry (potentially making separable structure appear continuous), such conclusions depend on the intrinsic dimensionality of the data and require robustness checks across embedding dimensionalities and hyperparameter settings, as performed in these studies (Kohli et al., 2025). This finding indicates that separable population readouts can arise from overlapping neuronal populations rather than from sharply segregated cell classes.

A similar conclusion was reached in dorsal premotor cortex during a vibrotactile categorization task. Diaz-deLeon et al. (2022) showed that categorical, temporal, and mixed responses occupied distinct population subspaces, yet individual neurons formed a continuous distribution in the embedding space, with no clear separation into discrete functional clusters (Fig. 4C, D). Thus, S2 and DPC are both examples in which separable population-level computations coexist with continuous single-neuron response profiles.

A complementary perspective comes from recurrent neural network models. Low-rank RNNs can generate multiple, approximately orthogonal readouts from a shared recurrent population without requiring anatomical segregation into discrete functional groups (Dubreuil et al., 2022). These models illustrate how structured low-dimensional geometry can arise from continuous mixed selectivity within a common circuit.

The near-orthogonality reported in these studies was quantified between demixed, task-marginalized subspaces that were validated across conditions and data splits, rather than between principal components; it is therefore not a byproduct of the orthogonality that PCA imposes by construction (Section 4). Converging evidence comes from methods that do not rely on variance-maximizing decompositions: single-trial flow-field inference revealed nearly orthogonal neural modes for evidence accumulation and decision commitment (Luo et al., 2025).

Together, these findings support a continuum view of cortical coding. Separable, and in some cases nearly orthogonal, population readouts should not be taken as evidence for a biological partition into separate circuits or cell classes. Conversely, the continuity of single-neuron responses does not weaken the relevance of separable population subspaces: because multiple signals coexist within the same population, separability may reduce interference between sensory, mnemonic, categorical, and decision-related readouts. The next sections apply this framework to temporal processing and then to its computational and translational extensions in artificial neural networks and brain–computer interfaces.

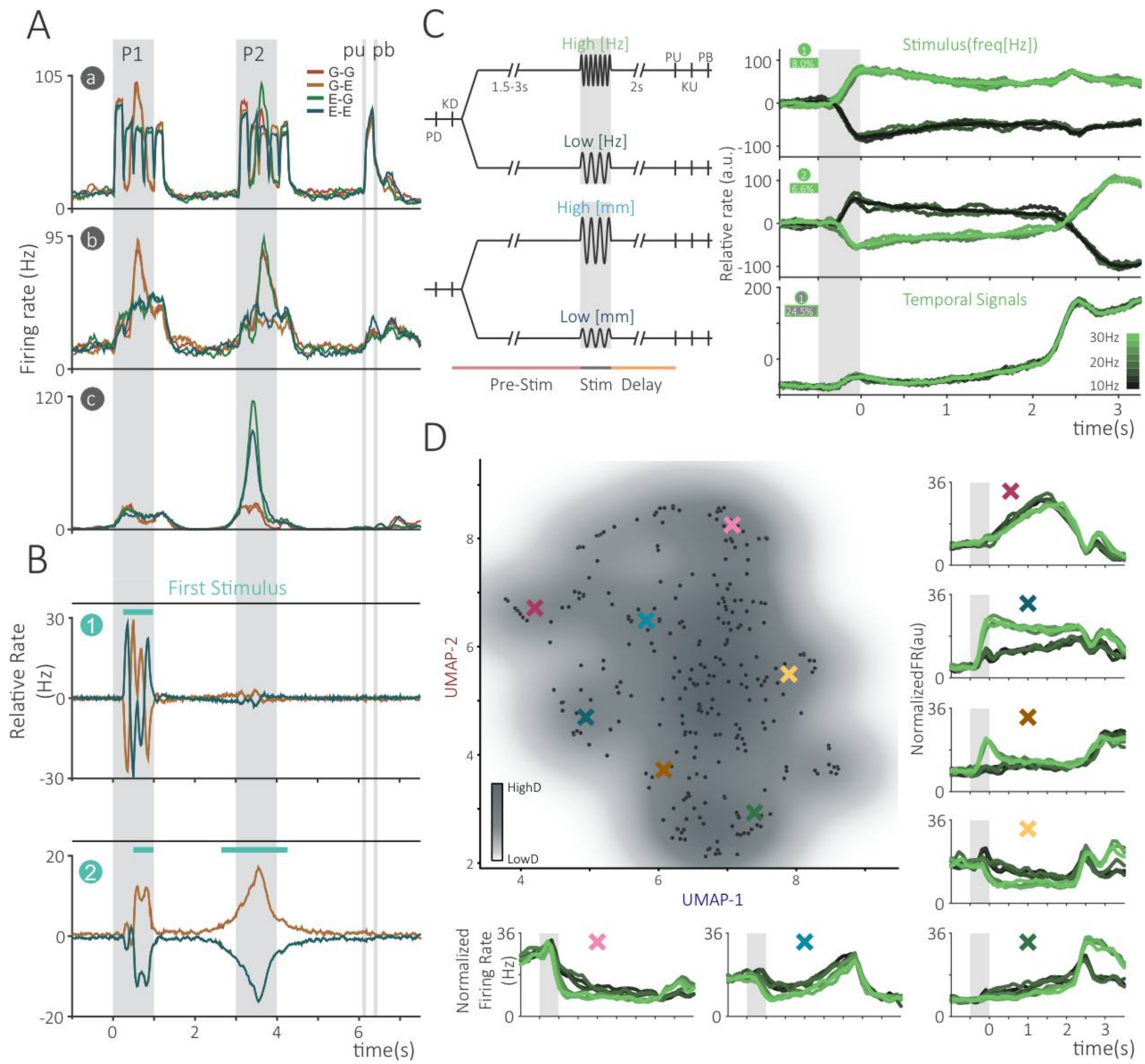

**Figure 4. Mixed selectivity, orthogonal subspaces, and continuity of single-neuron responses.** (A) Example S2 neurons showing firing rates during the two stimulation periods (P1, P2) across TPDT classes G–G, G–E, E–G, and E–E. (B) dPCA projections of S2 activity marginalized with respect to P1. Faithful sensory dynamics are expressed in the first component subspace, whereas transformed categorical dynamics are expressed in a second, nearly orthogonal subspace. (C) Vibrotactile categorization task using two stimulus features, flutter frequency (Hz) and amplitude (mm), each delivered at low or high values across pre-stimulus, stimulus, and delay epochs. Right: dPCA projections showing categorical components (top; frequency modality shown) and a temporal component (bottom), together with their explained variance. Traces are coloured by stimulus frequency level. (D) UMAP embedding of single-neuron response profiles showing heterogeneous but continuous response organization; coloured crosses mark representative functional profiles whose level-dependent trial-averaged responses are shown around the embedding. Panels A, B adapted from Bayones et al. (2024); C, D from Diaz-deLeon et al. (2022).

## 6. Temporal processing as flexible population dynamics

Because time is not specified by a static sensory feature but must be generated, maintained, compared, scaled, or transformed by neural dynamics, temporal processing is a useful case study for linking population geometry, circuit mechanisms, and behaviour.

Gámez et al. (2019) recorded from medial premotor cortex in macaques performing a synchronization–continuation task (Fig. 5A), in which the animal first tapped in phase with a visual metronome and then continued at the same rate without external pacing. Projections onto the second and third principal components traced a circular trajectory whose radius increased monotonically with the required interval, and the geometry of the loop was conserved between the synchronization and continuation phases (Fig. 5B). Thus, the amplitude of the population oscillation encoded the produced interval.

Wang et al. (2018) studied a complementary, cue-triggered form of interval production, executed by either a saccade or a key press (Fig. 5C). Single medial frontal cortex neurons exhibited temporal scaling: their firing-rate profiles stretched or compressed in proportion to the produced interval (Fig. 5D). At the population level, PC1 and PC2 preserved this scaling structure, whereas PC3 did not, placing scaling and non-scaling dynamics in approximately orthogonal population dimensions (Fig. 5E). Because PCA maximizes variance rather than temporal scaling, the authors introduced a scaling-component decomposition whose axes isolated the scaling subspace (Fig. 5F). In this basis, trajectories from different intervals overlapped when plotted against normalized time, indicating that interval differences were expressed primarily as changes in trajectory speed rather than trajectory shape. A recurrent neural network analysis provided a mechanistic interpretation: the cue set the initial and terminal fixed points, and its magnitude set the speed at which the state evolved between them, whereas recurrent dynamics determined the invariant trajectory along which this evolution occurred (Fig. 5G, H). Temporal scaling thus emerged from an approximately orthogonal factorization between an input subspace, which set fixed-point placement and speed, and a recurrent subspace, which generated the scaled trajectory.

Recent work in ventral premotor cortex (VPC) extends this population-dynamics view from interval production to temporal perception. Díaz et al. (2025) showed that temporal information can be reorganized according to behavioural context: the same premotor population supports more continuous, parametric dynamics when intervals must be compared, and more discrete, category-like dynamics when temporal information must be transformed into a categorical judgement. This result reinforces the idea that temporal processing is not encoded by a single clock-like variable, but by task-dependent population trajectories whose geometry reflects the required computation.

A further extension concerns how neural circuits incorporate the statistics of temporal events. Koppen et al. (2026) showed that cerebellar circuits learn prior probability distributions over temporal intervals during eyeblink conditioning, and that these temporal priors are reflected in Purkinje-cell simple and complex spike signalling, as well as in the low-dimensional dynamics of the recorded populations. Disrupting this activity abolished the predictive eyeblink responses, indicating that these cells are involved in generating prior-dependent behaviour. These findings show that neural activity can encode not only elapsed or produced time, but also the probabilistic structure of temporal events in the environment.

These studies show that temporal processing can be implemented through multiple forms of low-dimensional dynamics, including cyclic trajectories, temporal scaling, speed-controlled flow fields, context-dependent reconfiguration of population geometry, and prior-dependent temporal expectations.

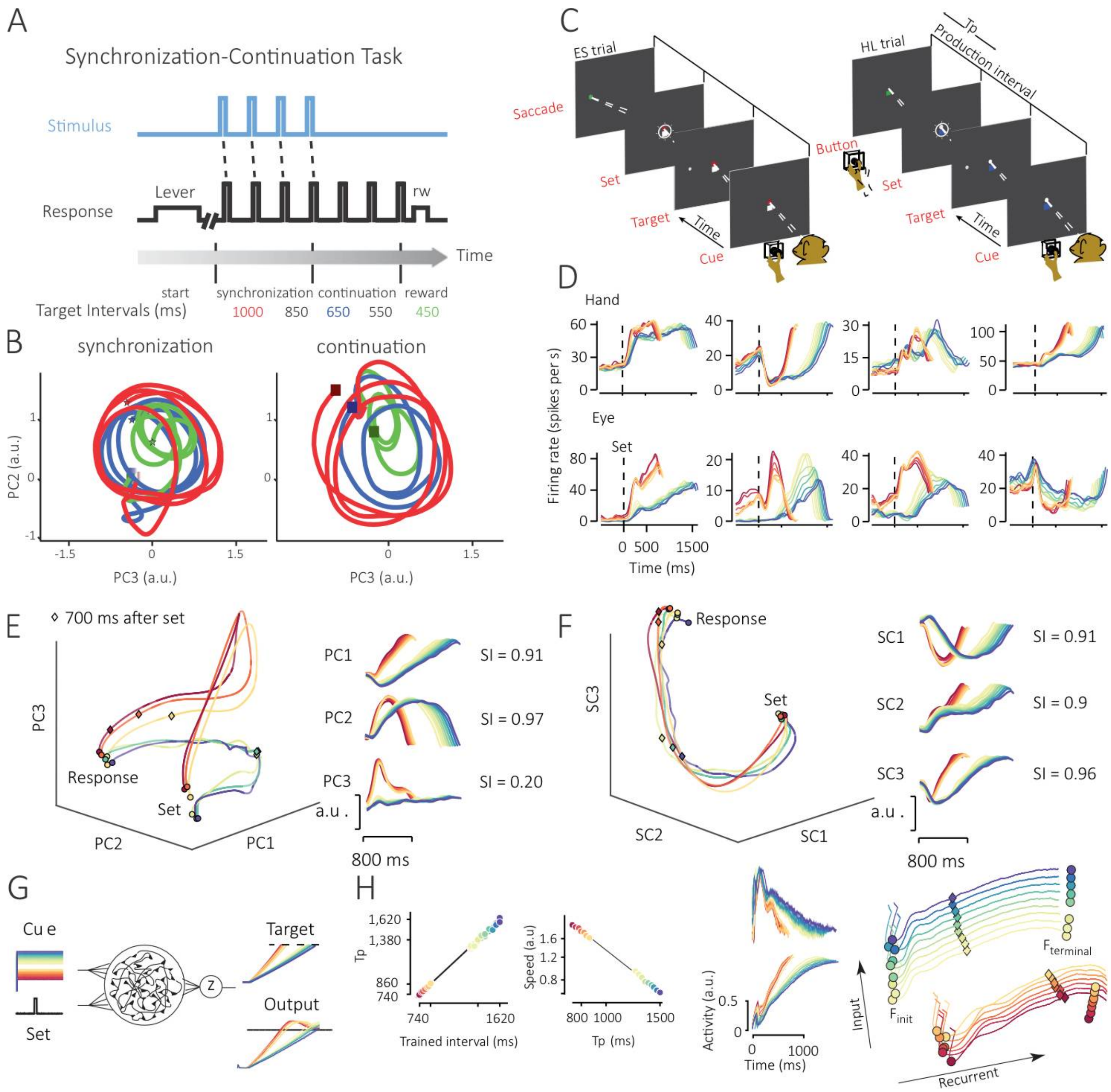

**Figure 5. Flexible temporal processing through cyclic and scaled low-dimensional dynamics.** (A) Synchronization–continuation task: the animal taps in synchrony with a periodic stimulus and then continues at the same tempo without external pacing, across target intervals of 450–1000 ms. (B) Population trajectories in the PC2–PC3 plane form cyclic dynamics whose geometry scales with the produced interval. (C) Set–Go interval-production task performed with eye or hand effectors: a cue and a Set pulse define the interval to be reproduced. (D) Example medial frontal cortex neurons showing temporally scaled activity across produced intervals. (E, F) Population trajectories from Set to response scale in time across intervals; the scaling index quantifies the degree of scaling for each component. (G, H) A recurrent network trained to produce the same intervals reproduces the behavioural scaling and provides a mechanistic account: the input sets the initial and terminal fixed points and the speed of evolution between them, whereas recurrent dynamics generate the invariant trajectory. Panels A, B adapted from Gámez et al. (2019), licensed under CC BY 4.0; C–H from Wang et al. (2018).

## 7. Single-trial population dynamics

The methods surveyed so far are usually applied to trial-averaged activity. This averaging improves signal-to-noise ratio and reveals condition-dependent structure, but it also removes trial-to-trial variability that can be informative about behaviour, internal state, noise correlations, and learning (Panzeri et al., 2022). Single-trial methods aim to recover this structure by estimating latent population dynamics on individual trials rather than only in condition averages.

Several families of methods have become central to this problem. ***Tensor component analysis (TCA)*** operates on a neuron × time × trial tensor, rather than on the condition-averaged tensor introduced in Section 3, and factorizes it into low-dimensional components with neuron, temporal, and trial factors (Williams et al., 2018). Because it does not require predefined task labels, TCA is useful for exploratory analyses: trial factors can subsequently be related to behavioural variables, latent states, learning, or temporal covariates. ***Sparse component analysis (SCA)*** is likewise label-free and, applied to single-trial recordings, can isolate temporally sparse factors associated with distinct computations (Zimnik et al., 2026). ***Model-based targeted dimensionality reduction (mTDR)*** takes a more supervised approach, defining low-dimensional components associated with specified task variables within a probabilistic regression framework (Aoi and Pillow, 2018; Aoi et al., 2020). This makes it suited to cases in which the relevant covariates are known and the goal is to isolate the population dimensions that correspond to each of them.

The ***generalized multilinear model (GMLM)*** extends this logic by combining generalized linear modelling with tensor factorization, so that population activity can be related to richer sets of task and behavioural covariates through low-rank structure (Latimer and Freedman, 2023). This is valuable in flexible or naturalistic tasks in which events, choices, and behavioural variables cannot be reduced to a small number of discrete conditions. ***Latent factor analysis via dynamical systems (LFADS)*** takes a different approach: it is a sequential autoencoder based on recurrent neural networks that infers smooth latent dynamics and denoised firing rates from single-trial spiking activity (Pandarinath et al., 2018). LFADS therefore bridges dimensionality reduction, dynamical-systems modelling, and nonlinear generative modelling.

A related class of single-trial dynamical models infers latent trajectories together with the dynamics that generate them and the geometry through which they are expressed in heterogeneous single-neuron responses. Genkin et al. (2025) developed such an approach and applied it to recordings from primate premotor cortex, simultaneously inferring a low-dimensional decision variable and the nonlinear tuning functions of individual neurons to that latent state. The analysis showed that heterogeneous firing-rate profiles could arise from diverse tuning to a common dynamic choice variable, whose inferred dynamics were consistent with an attractor mechanism for decision formation.

More recent flow-field inference approaches go a step further and estimate how latent states evolve during behaviour. Luo et al. (2025) applied one such method, ***flow-field inference from neural data using deep recurrent networks (FINDR)***, to infer single-trial decision dynamics from frontal-cortex and striatal spiking activity during an auditory evidence-accumulation task. This analysis revealed a rapid transition from an input-driven evidence-accumulation regime to a more autonomous decision-commitment regime, accompanied by a change in neural mode: the flow direction during commitment was largely orthogonal to that during evidence accumulation. These results therefore link single-trial latent-variable modelling to the orthogonality framework described above and show that single-trial dynamics can reveal both where population activity lies in state space and how the underlying dynamical regime changes during behaviour.

These methods occupy different points along a spectrum. TCA and SCA are useful when the structure of trial-to-trial variability is not known in advance. mTDR and GMLM are better suited to hypothesis-driven analyses in which behavioural or task variables are explicitly modelled. LFADS is the natural choice when the goal is to infer latent dynamics and denoise single-trial spike trains, especially in datasets with enough neurons and trials to support nonlinear generative models. Single-trial dynamical inference methods, such as those used by Genkin et al. (2025) and Luo et al. (2025), add a further layer by estimating the latent dynamics, representational geometry, and regime changes underlying decision formation. In practice, these approaches are complementary: applying more than one method to the same dataset can help distinguish robust population structure from method-specific features (Valente et al., 2021; Umakantha et al., 2021; Boucher et al., 2023; Genkin et al., 2025; Luo et al., 2025).

## 8. Artificial neural networks as mechanistic hypotheses

Artificial neural networks make it possible to formulate and test mechanistic hypotheses about cortical computation. They allow one to specify a task, train a circuit-like model, and then analyse whether the resulting dynamics resemble those observed in neural populations (Song et al., 2016; Yang et al., 2019; Yang and Wang, 2020). For population dynamics, their value goes beyond reproducing behaviour: they expose candidate mechanisms, such as fixed points, attractors, low-dimensional manifolds, input–output rotations, or subspace factorization, that can then be compared with biological data.

***Task-trained RNNs with biological constraints.*** Recurrent neural networks that incorporate constraints such as Dale’s law, excitatory–inhibitory structure, or sparse connectivity can reproduce heterogeneous single-neuron responses while generating structured population dynamics similar to those observed in frontal and motor circuits (Song et al., 2016; Russo et al., 2018; Yang et al., 2019). The temporal-scaling model discussed in Section 6 is a clear example: after training, the network revealed a fixed-point structure in which the input set the initial and terminal fixed points and the speed of evolution between them, whereas recurrent dynamics generated the invariant trajectory (Wang et al., 2018). Thus, biological constraints restrict the solution space and can also make learned solutions more interpretable as candidate circuit mechanisms.

***Reservoir computing and trained readouts.*** Reservoir computing is a complementary modelling strategy in which recurrent connectivity is kept fixed, or only weakly modified, and task learning is concentrated in readout or feedback weights. In echo-state networks and liquid-state machines, a high-dimensional recurrent reservoir transforms inputs into rich temporal activity patterns, from which task-specific variables can be decoded by trained outputs (Jaeger, 2001; Maass et al., 2002). FORCE learning extends this idea by using rapid recursive least-squares updates to generate stable temporal patterns from initially complex recurrent activity (Sussillo and Abbott, 2009; Laje and Buonomano, 2013). This architecture is relevant for population dynamics because it separates a shared recurrent substrate from task-specific readout geometry: the same reservoir can support different computations through different output projections. Reservoir-based models are therefore a useful framework for studying cognitive flexibility, multitask computation, and the stability or reconfiguration of population manifolds across contexts.

***Low-rank RNNs.*** Another class of models makes the link between connectivity and geometry more explicit. In low-rank RNNs, the recurrent connectivity matrix $W$, which defines the strength of connections among $N$ units, is factorized as $W = UV^T$, with rank $R \ll N$. Here, $U$ and $V$ contain $R$ connectivity-defined patterns that constrain network activity to evolve within a low-dimensional dynamical subspace (Mastrogiuseppe and Ostojic, 2018; Dubreuil et al., 2022). This framework is a mechanistic counterpart to empirical low-dimensional structure: instead of discovering such structure after recording the population, the model specifies how low-dimensional dynamics can arise from structured recurrent connectivity. Low-rank RNNs have been used to explain context-dependent selection, flexible temporal processing, and the coexistence of multiple readouts in a shared recurrent population (Dubreuil et al., 2022; Beiran et al., 2023). They are also analytically tractable, because the effective dimensionality of the dynamics is constrained by the rank of the connectivity rather than discovered only post hoc.

***Task-trained RNNs as dataset-specific models.*** RNNs can also be used as models of specific neural datasets. In VPC, population recordings during bimodal detection revealed separable dynamics for tactile, acoustic, and stimulus-absent conditions, together with delay-period activity related to maintenance and behavioural report (Andrade-Ortega et al., 2025). The same study trained RNNs on the task and showed that similar low-dimensional trajectories could emerge from recurrent circuit dynamics. This is a concrete bridge between descriptive population geometry and mechanistic modelling: the RNN is not treated as a black-box decoder, but as a dynamical system whose internal states can be analysed with the same tools used for neural populations.

***Pretrained models and representation learning for neural data.*** A final development concerns models that learn neural representations across datasets rather than within a single session. These methods differ in kind and should not be grouped under a single label. CEBRA is a contrastive representation-learning method that embeds neural activity, optionally jointly with behavioural labels, into a low-dimensional latent space whose geometry can be compared across sessions and animals (Schneider et al., 2023). The Neural Data Transformer (NDT) is a transformer sequence model that infers latent firing rates from spiking activity within a dataset (Ye and Pandarinath, 2021), and later versions extend it with pretraining across sessions and subjects (Ye et al., 2023). POYO comes closest to what machine learning calls a foundation model (a large model pretrained on broad data and then adapted to many downstream tasks) by tokenizing individual spikes from heterogeneous recordings spanning sessions, animals, and tasks and pretraining a single transformer that transfers to new datasets with limited fine-tuning (Azabou et al., 2023). These approaches extend the logic of LFADS from single-dataset latent-dynamics inference toward more general-purpose neural representations (Pandarinath et al., 2018). Their relevance for population dynamics is twofold: they can recover low-dimensional latent structure from complex recordings, and they provide practical tools for chronic datasets in which neural identity, tuning, and recording conditions drift across days.

These approaches position artificial neural networks as instruments for generating and testing mechanistic hypotheses, rather than merely as predictive models. Task-trained RNNs test whether a proposed circuit architecture can solve the relevant behaviour and produce neural-like population dynamics. Reservoir models show how a shared recurrent substrate can support flexible readouts. Low-rank RNNs reveal how structured connectivity can generate low-dimensional dynamics. Pretrained and representation-learning models yield latent representations that transfer across heterogeneous recordings. Across these cases, the central criterion is whether the model, beyond reproducing behaviour, recapitulates the geometry, timescales, and task-dependent transformations observed in neural populations.

## 9. From population dynamics to brain–computer interfaces

The tools developed throughout this review (low-dimensional manifolds, orthogonal subspaces, demixed latent factors, and single-trial latent dynamics) were originally introduced to interpret cortical activity. A parallel line of work has treated the same structures as engineering primitives: if cognitive, sensory, or motor variables occupy structured subspaces, they can be decoded in real time and routed to external effectors or feedback channels. This is the operating principle of brain–computer interfaces (BCIs), particularly intracortical BCIs, which use neural population activity to restore movement, somatosensory feedback, handwriting, or speech in people with paralysis. Figure 6 traces this arc: from recurrent-network models that formalize how low-dimensional computations arise from connectivity (Fig. 6A, B), through models pretrained across datasets that learn representations shared across animals, sessions, and tasks (Fig. 6C), to BCIs that translate population activity into movement and communication (Fig. 6D–F).

***Motor BCIs and the low-dimensional control space.*** Early intracortical BCIs decoded movement direction, hand velocity, or endpoint position from motor-cortex population activity, building directly on population-vector and state-space views of motor control (Georgopoulos et al., 1986). Modern systems, including BrainGate and related platforms, exploit the fact that task-relevant motor activity often occupies a low-dimensional subspace far smaller than the number of recorded units, typically on the order of a few to a dozen dimensions (Sadtler et al., 2014; Gallego et al., 2017). Decoding can therefore be framed as estimating the user's trajectory within this low-dimensional control space (Fig. 6D). Linear decoders such as Kalman filters perform well in this regime, while recurrent and nonlinear architectures can improve robustness by denoising single-trial dynamics and exploiting temporal structure (Pandarinath et al., 2018). A related principle is that preparatory and movement-related activity occupy partially separate subspaces: readouts can be designed to avoid preparatory null dimensions, reducing the risk that motor planning is translated into unintended movement (Kaufman et al., 2014; Elsayed et al., 2016).

***BCI learning as a causal test of population geometry.*** Closed-loop BCIs make it possible to perturb the relationship between neural activity and behaviour directly. Sadtler et al. (2014) showed that monkeys learned BCI perturbations more readily when the required neural activity patterns remained within the intrinsic population manifold (a linear subspace estimated by factor analysis) than when the decoder required outside-manifold patterns. This result gave the low-dimensional description causal force: population geometry does not merely describe neural activity; it constrains which patterns are immediately learnable. Subsequent work refined this picture. Golub et al. (2018) found that learning can proceed through neural reassociation, in which existing activity patterns are remapped onto new behavioural outputs. Hennig et al. (2018) used BCI perturbations to define redundant activity patterns and found that this redundancy is structured rather than arbitrary. Oby et al. (2019) then demonstrated that, over longer learning periods, monkeys can generate new neural activity patterns that support new BCI skills. These studies show that motor-cortical population structure is both constraining and plastic: it defines the initial repertoire of accessible patterns, but long-term learning can expand or reshape that repertoire.

***Stable population dynamics enable stable decoders.*** A major obstacle for chronic BCIs is representational drift: the recorded neurons and their apparent tuning can change across days, even when behaviour remains stable. Population geometry suggests a way around this problem. Gallego et al. (2020) showed that the low-dimensional latent dynamics of motor cortex can remain stable across weeks and months despite turnover in the recorded neurons (Fig. 6E). Safaie et al. (2023) found preserved latent dynamics across animals performing similar behaviours, and Rule et al. (2020) showed that stable task information can be extracted from unstable neural ensembles. These results motivate decoder-stabilization approaches that align the low-dimensional spaces of neural activity recorded on different days rather than re-estimating single-neuron tuning from scratch. Degenhart et al. (2020) demonstrated this directly by stabilizing BCI control through such an alignment across recording instabilities. The BCI literature often refers to this phenomenon as manifold stability. In the terminology adopted here, what these studies establish is the stability of low-dimensional latent dynamics and of the subspace that contains them, which is necessary, but not by itself sufficient, to identify a manifold.

***Bidirectional and communication BCIs.*** Population dynamics also matter for closed-loop sensory feedback. Motor control is not only a decoding problem: natural movement depends on sensory consequences that update ongoing action. Bidirectional BCIs address this by combining motor decoding with intracortical microstimulation of somatosensory cortex. Flesher et al. (2021) showed that stimulation-evoked tactile sensations improved robotic arm control in a person with tetraplegia. These results take the population-dynamics framework from readout to closed-loop sensorimotor control. In parallel, communication BCIs have carried the state-space approach from arm and cursor control to handwriting and speech. Willett et al. (2021) decoded attempted handwriting from motor cortex by exploiting the fact that intended characters follow characteristic neural trajectories rather than static activity patterns (Fig. 6F). Subsequent work decoded attempted speech from motor and speech-related cortical areas, using intracortical or electrocorticographic recordings combined with modern sequence decoders and language models (Willett et al., 2023; Metzger et al., 2023; Card et al., 2024).

Together, these studies show that BCIs are not simply applications of population neuroscience. They are also experimental platforms for testing the causal role of manifolds, subspaces, and latent dynamics. In motor BCIs, low-dimensional subspaces define the control space. In BCI learning, the intrinsic manifold constrains which neural patterns are easy or difficult to acquire. In chronic decoding, subspace alignment stabilizes control despite neuronal drift. In bidirectional systems, motor and sensory signals must be coordinated within a closed-loop state space. In communication BCIs, latent trajectories support rapid decoding of handwriting and speech. Population geometry therefore provides both the conceptual language and the engineering substrate for modern neurotechnology.

***Open challenges.*** Three directions define the current frontier. First, closed-loop co-adaptation remains poorly understood. During BCI use, both the decoder and the user's cortical activity change, and the geometry of this joint adaptation is still only partially characterized. Work on intrinsic manifolds and BCI learning shows that population geometry constrains which activity patterns are readily accessible, but also that long-term learning can expand or reorganize this repertoire (Sadtler et al., 2014; Oby et al., 2019). Second, generalization across tasks, effectors, and recording sessions is a central challenge. A BCI trained for one behaviour should ideally transfer to related behaviours with minimal recalibration, but the extent of this transfer will depend on how similar the underlying population subspaces, effectors, and task demands are across behaviours. This question is beginning to be probed directly with BCI paradigms: learning a new decoder mapping leaves a memory trace in motor cortex that coexists with proficient performance of a familiar mapping (Losey et al., 2024), and the temporal structure of population activity imposes dynamical constraints on which new activity patterns can be produced (Oby et al., 2025). At the same time, the low-dimensional control spaces used by current decoders may underestimate the degrees of freedom available to motor cortex: population responses can support flexible control of individual motor units beyond a single common drive (Marshall et al., 2022); dexterous BCIs may therefore need to access higher-dimensional components of population activity. Models pretrained across datasets, such as POYO (Azabou et al., 2023), provide one promising route because they learn latent structure from heterogeneous recordings and can be fine-tuned to new datasets or behaviours. Third, bidirectional interfaces that combine decoding with patterned stimulation (Flesher et al., 2021) require a deeper understanding of how motor commands and sensory feedback can coexist in shared neural state spaces. In this setting, the demixing and orthogonality principles discussed above may help design feedback channels that minimize interference with decoded output.

The convergence documented in this review is therefore both conceptual and practical. The same population-geometry framework that explains working memory, decision-making, and temporal processing now guides the design of neuroprosthetic systems intended to restore movement, communication, and autonomy.

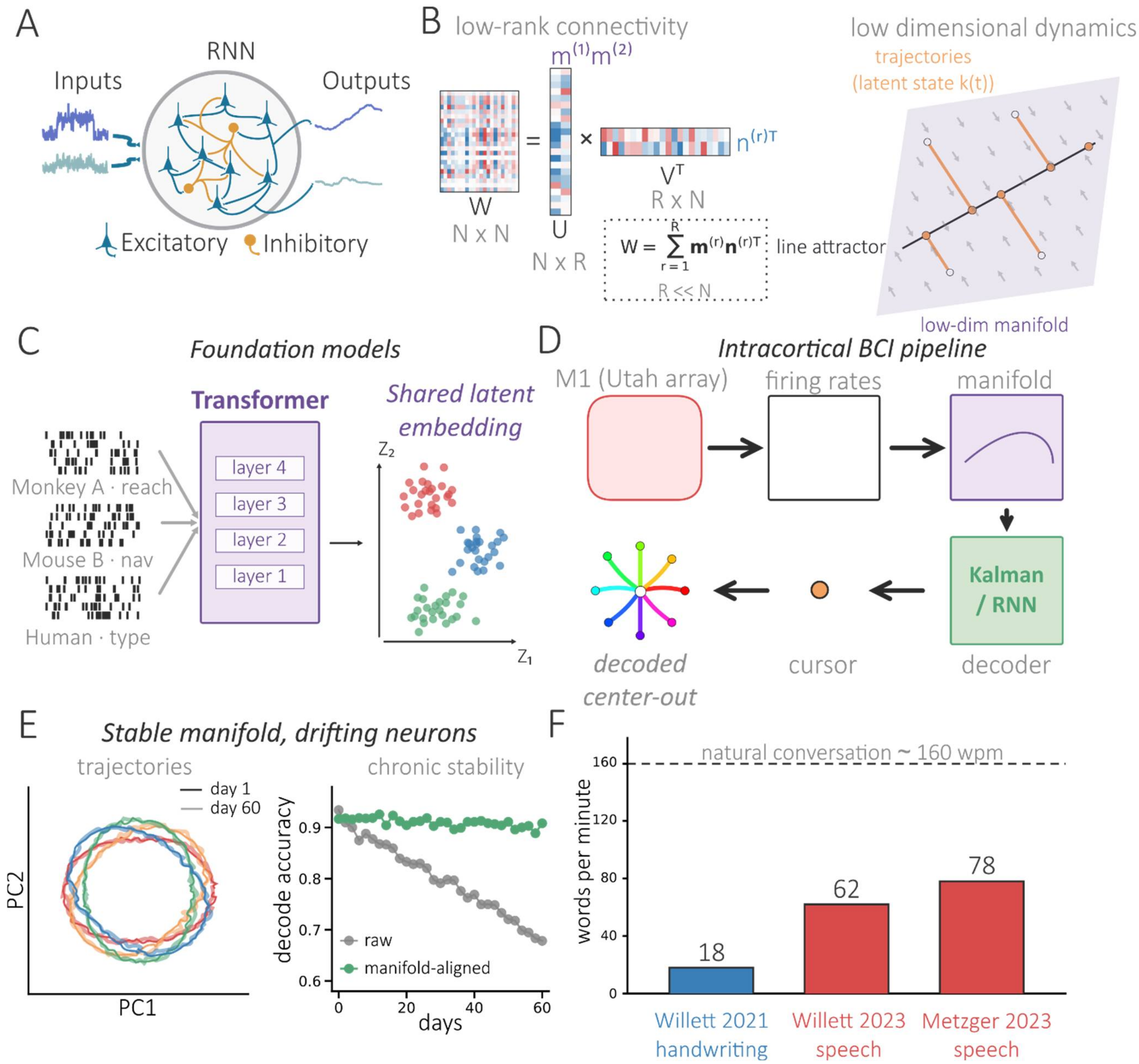

**Figure 6. From population dynamics to artificial networks and restorative neurotechnology**. (A) Recurrent networks with separate excitatory and inhibitory units are trainable mechanistic models of population computation. (B) Low-rank connectivity, $\boldsymbol{W} = \boldsymbol{U}\boldsymbol{V}^{\boldsymbol{T}} = \sum_{r=1}^{R} \boldsymbol{m}^{(r)}\boldsymbol{n}^{(r)T}$ with R ≪ N constrains activity to a low-dimensional subspace within which latent dynamics, here illustrated as a line attractor, can implement computations such as parametric memory. (C) Pretrained models for neural data: spikes from heterogeneous recordings spanning species, tasks, and sessions are tokenized and passed through a transformer to learn a shared latent embedding that transfers across datasets. (D) Intracortical BCI pipeline: motor-cortical activity recorded with a Utah array is projected onto a low-dimensional subspace and mapped by a decoder, such as a Kalman filter or RNN, to drive an external effector, here illustrated as a centre-out cursor task. (E) Low-dimensional latent dynamics can remain stable over time even as recorded units drift: aligning the low-dimensional spaces recorded on different days maintains decoding performance more effectively than a fixed decoder. (F) Communication-BCI performance has advanced rapidly toward naturalistic communication rates, from attempted handwriting to attempted speech decoding, with representative online rates increasing from ~18 to ~62–78 words per minute in recent systems. Panels A–D and F are original schematics; panel E is based on Gallego et al. (2020) and Safaie et al. (2023).

## 10. Conclusions

This review traced a progression from single-neuron coding to population geometry, computation, and neurotechnology. Single-unit analyses remain indispensable for characterizing selectivity, latency, and heterogeneity, but population-level methods reveal structure that is not visible neuron by neuron: low-dimensional subspaces, separable readout axes, dynamic trajectories, and task-dependent geometric transformations. Across domains, the evidence reviewed here converges on two principles. Cognitive variables can occupy separable or near-orthogonal population dimensions, which limits interference between sensory, mnemonic, temporal, decision-related, and motor signals. This separation does not require discrete cell classes: in somatosensory and premotor cortex, it emerges from continuous, mixed-selective populations sharing a common substrate. The population-dynamics framework is productive because it operates at several levels. It describes data through dimensionality reduction and geometry. It suggests mechanisms through attractors, flow fields, and structured connectivity. It tests these mechanisms through artificial networks trained on matched tasks. And in brain–computer interfaces, it becomes operational: population geometry can be decoded, stabilized, perturbed, and used to restore movement or communication.

Important gaps remain. The relationship between anatomical connectivity and dynamical geometry is still an open question (Wang, 2022). Biologically realistic plasticity in artificial models is an unsolved problem (Zenke et al., 2017). The continuity of single-neuron responses raises unresolved questions about how functional classes should be defined (Rossi-Pool et al., 2021; Bayones et al., 2024). Finally, population-level conclusions depend on analytical choices whose assumptions must be validated rather than taken for granted. These challenges define a set of tractable questions for future work: how biological circuits generate, maintain, and modify the population geometries that underlie cognition and neurotechnology.

**CRediT authorship contribution statement:** Conceptualization: R.R.-P. Writing – original draft: L.B., D.N., R.R.-P. Writing – review & editing: L.B., D.N., R.R.-P. Visualization: L.B. Supervision: R.R.-P.

**Declaration of competing interest:** The authors declare no competing interests.

**Acknowledgements:** During the preparation of this work the authors used Claude (Anthropic) to edit language and check consistency across the manuscript. After using this tool, the authors reviewed and edited the content as needed and take full responsibility for the content of the published article.

**Funding:** This work was supported by grant PAPIIT-IN203825 from the Dirección General de Asuntos del Personal Académico of the Universidad Nacional Autónoma de México, and by grant SECIHTI-CBF-2025-I-2054 from the Secretaría de Ciencia, Humanidades, Tecnología e Innovación (to R.R.-P.).

**Data availability:** No new data were generated or analysed in this review.